\documentstyle[12pt,aasms]{article}
\begin{document}
\begin{center}
\section*{\bf Surface Quantum Effects in a Fireball Model of Gamma Ray
Bursts}

\vspace{5mm}
R. Lieu$^1$, Y. Takahashi$^1$, T.W.B. Kibble$^2$, J. van Paradijs$^{1,3}$, and
A.G. Emslie$^{1}$

\vspace{1.5mm}

$^1$ Department of Physics, University of Alabama, Huntsville, AL 35899. \\
%(lieur@cspar.uah.edu, yoshi@cosmic.uah.edu) \\
$^2$ Blackett Laboratory, Imperial College, London SW7 2BZ. \\
%(t.kibble@ic.ac.uk) \\
$^3$ Astronomical Institute `Anton Pannekoek', 
University of Amsterdam, 1098 SJ Amsterdam, The Netherlands. \\

\end{center}

\vspace{2mm}

{\it Recent rapid advances in the observations of Gamma Ray
Bursts (GRBs) lend much confidence to the 
cosmological fireball (FB) model: there is now compelling evidence that the
radiation is emitted by a relativistic shock
where a high speed upstream flow terminates.
The question concerning what generates such an
outflow is central to our search for the 
ultimate trigger mechanism.  A key
requirement, not well
explained by current theories, is that the flow must have
high entropy-to-baryon ratio.  In this Letter we point out that a
quantum discharge induced by 
the radiation of the initial FB
may be the explanation.  The effect is likely to be relevant, because
the FB energy density inferred from  GRB data is large enough that
the radiation pressure
leads to the formation of a surface electric field which is
unstable to pair creation.  Under suitable conditions,
such as those of a supernova core, the discharge can convert
a substantial fraction of the FB energy into 
surface pairs.  This pair plasma is not
contaminated by FB baryons because it is formed outside the FB.
We demonstrate that the pairs can then develop 
relativistic bulk expansion, reaching a maximum speed that meets the
constraints required to form a GRB.}

\vspace{2.5mm}

The discovery of GRB
afterglows at X-ray, optical and radio wavelengths (Costa et al 1997,
van Paradijs et al 1997, Frail et al 1997)
quickly led to a final settlement of the issue concerning
distance scale: the high redshift associations of GRB-970508
(Metzger et al 1997) and
GRB-971214 (Kulkarni et al 1998)
indicate that
GRB must be of cosmological origin, and that 
its observed fluence 
corresponds to a total energy release 
$\epsilon_o \geq$ 10$^{52-54}$ ergs.
In this paper
we take $\epsilon_o \sim 4 \times 10^{52}$ ergs as representative.
Since such a
value suggests
that a process at the end point of stellar
evolution  (e.g., merging of two neutron stars,
collapse of a massive star at its evolutionary end point)
is responsible for triggering
the GRB, it is reasonable to assume that the radius 
of the trigger is $r_o \sim$ 10 km.
To the
lowest order, therefore, one may consider the uniform and simultaneous 
release of
energy $\epsilon_o$ over a radius $r_o$, and
deduce an initial energy
density $U_o \sim 10^{34}$ ergs cm$^{-3}$
(hereafter $U_o$ will be expressed in such a unit,
i.e., $U_{34}$).  Any internal spatial gradients will after all be rapidly
smoothed out in the course of the expansion.

The classic difficulty in explaining a GRB lies not with the availability of
a violent energy release mechanism, but 
rather with our ability to account for
the basic observed properties
of the burst.  A FB of
energy density $U_{34} = 1$ is extremely optically thick to
Compton-related quantum processes, so that the thermalization
timescale is very short.  The plasma explosion
(i.e. $\gamma$, e$^+$e$^-$, and baryons) 
is therefore unobservable electromagnetically, and when the
plasma expands outwards to become optically thin,
its flow turns relativistic {\it in situ}.
The observed 
non-thermal spectrum and the long
duration (1 - 10 s) of a GRB implies that most of the energy we detect is
from the bulk motion, not the radiation, of the outflow (Meszaros and
Rees 1993, Paczynski 1993).
The conversion 
back to
random energy takes place via the extremely rapid
particle acceleration at a relativistic
shock (Quenby and Lieu 1989)
when the flow is
finally terminated by the ambient medium, and
the energy is then radiated away more
leisurely.  This notion of `delayed emission' 
was predicted (Meszaros \& Rees 1997, Paczynski and Rhoads 1993, Vietri 1997)
and is confirmed by the discoveries
of afterglows at longer wavelengths.

The evolution of the FB from creation to observability is, however,
easily jeopardized by the presence of trace baryons in the flow, which
drastically slow down the bulk motion.  Indeed, 
canonical FB models (Meszaros and Rees 1993) presume the mass $M_b$
of baryons in the FB obeys
$\xi = M_b/M_\odot < 10^{-4} = \xi_c$ (resulting in an outflow that reaches
maximum bulk Lorentz
factor of a few $\times$ 10$^2$)
although hitherto there does not exist
a satisfactory explanation of how the
requirement may be met.
This {\it Letter} addresses a new
physical mechanism which may provide vital clues.

It has long been recognized (Schwinger 1951) that when a static
electric field exceeds a critical value 
$E = m_e^2 c^3/e \hbar = E_c$, corresponding to an 
electron acceleration of
a$_c \sim$ 2.4 $\times$ 10$^{31}$ cm s$^{-2}$, it is unstable with
respect to pair production.  Under the circumstance, a virtual
e$^+$e$^-$ pair is accelerated
in opposite directions
by the electric field to the speed of light within a Compton
wavelength.  This greatly reduces the probability of annihilation:
real pair creation may then take place at the energy expense of the
field.  The process is analogous to a lightning discharge, except that
it happens in vacuum.
The rate of pair production per unit volume is given by the formula
Schwinger derived:
\begin{equation}
\frac{d^2N}{dVdt} = \frac{\alpha^2 E^2}{\pi^2 \hbar}
\sum_{m=1}^{\infty} m^{-2} exp 
\left(\frac{-m \pi E_c}{E} \right)~~~~{\rm cm}^{-3}~{\rm s}^{-1}
\end{equation}
where $\alpha$ is the fine
structure constant and $N$ is the number of pairs created.  Using
this formula, we find that
for $E \sim E_c$ the `vacuum breakdown' causes the field to dissipate
its energy in a timescale of $\leq$ 10$^{-17}$ seconds, resulting
in a gamma ray and pair plasma.

Owing to the strength of $E_c$, this instability has not been realized
in the laboratory.  However, in the context of a GRB, a short-term but
intense surface
electric field
may develop as a result of the large radiation 
pressure of the explosion.
Consider a FB
of energy density $U$ which
expands into its immediately vicinity at speed c, the
latter having a
proton number density $n_i$ before the disturbance.  
Assuming that the fraction of $U$
due to radiation (i.e. $\gamma$, e$^+$e$^-$) is $\sim$ 100 \%,
the Eddington
force on an electron is
$\sim \sigma_{c} U$ ($\sigma_c$ being the Compton cross
section).  Although in an extreme
Compton limit the prevalence of forward scattering 
tends to undermine this force, the Eddington value remains correct if the
mean free path is short enough that
repeated Compton events can `re-use' the scattered 
radiation, which is likely to be
the situation just outside a GRB FB.
The radiation
will accelerate the electron
away from its neighbouring protons until a counteracting electric field
develops between them.  
At this
point the separation distance $\Delta r$ reaches a value $l$  given by:
\begin{equation}
\sigma_c U = 4 \pi n_i e^2 l
\end{equation}

Non-relativistically the
electrons continue to 
surge forward until $\Delta r$ is at the maximum value of
$2l$,
and subsequently
$\Delta r$ oscillates between zero and maximum at the plasma
frequency $\omega_p = \sqrt{4 \pi n_i e^2/m_e}$ 
while the two layers of charges
accelerate outwards.  The `equilibrium separation' $l$ as depicted in (2)
is dependent on $n_i$.  Now the baryon density of the stationary medium at the
surface of FB is unlikely to be orders of magnitude smaller
than that inside the FB.
Assuming that 
the density ratio is $\eta \sim$ 1 - 10 \%, and that the 
FB density within is initially
$\sim$ a few $\times$ 10$^{36-38}$ cm$^{-3}$ (i.e.
$\xi \sim$ 0.01 - 0.5 for $r_o =$ 10 km),
we shall express $n_i$ in units of
10$^{36}$ cm$^{-3}$, i.e. $n_{36}$.
Since the radiation blast
comprises photons of energy $kT \gg m_e c^2$ 
where $aT^4 = U$, the high energy limit of the Compton cross
section may be used:
\begin{equation}
\sigma_c = 8.8 \times 
10^{-27} U_{34}^{-\frac{1}{4}} f(U_{34})~~~~{\rm cm}^2
\end{equation}
where 
\begin{equation}
f(U_{34}) = 1 + 0.04~{\rm ln}~U_{34}
\end{equation}
$l$ in (2) is then given explicitly by:
\begin{equation}
l \sim 3 \times 10^{-11} n_{36}^{-1} U_{34}^{\frac{3}{4}} f(U_{34})~~~~
{\rm cm}
\end{equation}

Note that because in the case of an initial GRB FB, $U$ is
large enough to induce
`super-Schwinger' acceleration $a \gg a_c$ (i.e. $U_{34} \geq 1$), the
electron will reach speed $c$ having only 
traversed a small fraction of $l$, and
relativistic corrections apply to most of the motion.
Nonetheless
there remains the
oscillatory nature of $\Delta r$, now defined as the charge separation
measured with respect to our laboratory system $\Sigma$, so
pertinent results may
still be estimated.  For instance, at maximum $\Delta r$ the electrons are
at rest with respect to $\Sigma$, and the radiation and electric
forces are once again given by $\sigma_c U$ and
$4 \pi n_i e^2 \Delta r$
respectively, with the latter having {\it exceeded} the former.  Thus $l$ as
given by (5)
provides a lower limit to the
growth of the electric field.
The field will oscillate between
$E = 0$ and $E \geq \sigma_c U/e$
as the separation oscillates between $\Delta r = 0$
and $\Delta r \geq l$.

If $\sigma_c U/e > E_c$
the field can, in
principle, discharge into pairs once the separation $\Delta r$ is
large enough that $E \sim E_c$.
In reality, however, timescale comparisons indicate that the
field can reach peak strength, with discharge happening 
efficiently at this point only if
conditions are favorable (see below).
Microscopically the pair production process returns an electron
to the proton side of the double-layer and  creates a positron on the
electron side
(see Figure 1).
The pairs are free to escape, as they are created
{\it outside} the FB, and as
as we ignore the
effects of a magnetic field in this first attempt on the problem.
Moreover, the discharge is sustained at the FB surface so long as a
super-Schwinger surface field continues to be regenerated by the radiation
pressure of a homogeneously expanding FB.

To further investigate the Schwinger mechanism,
equation (1) may be expressed in energy units since
the temperature $T$ of the post-discharge e$^+$e$^-$ pairs is
obtainable by equating the energy density at peak field
\begin{equation}
\frac{E^2}{8 \pi} \sim \frac{1}{8 \pi} \left(\frac{\sigma_c U}{e} \right)^2
\end{equation}
with $aT^4$.
The mean energy of each pair, $3kT$ ($\gg m_e c^2$), 
is then multiplied by (1) to give a
peak volume energy loss rate due to discharge:
\begin{equation}
\frac{d^2 \epsilon}{dVdt} = 4.7 \times 10^{52} U_{34}^{\frac{15}{8}} 
[f(U_{34})]^{\frac{5}{2}} g(U_{34})~~~~{\rm ergs}~{\rm cm}^{-3}~{\rm s}^{-1}
\end{equation}
where
\begin{equation}
g(U_{34}) = \sum_{m=1}^{\infty} m^{-2}
exp \left\{- 7.51 \times 10^{-4} U_{34}^{-\frac{3}{4}} 
[f(U_{34})]^{-1} m \right\} \nonumber
\end{equation}
is $\sim 1$ for $U_{34} \geq 10^{-4}$ and $\ll 1$ otherwise (the latter
means $E < E_c$, i.e.
the low radiation pressure only induces a
sub-Schwinger surface field, so no discharge takes place).  A simple
division of (6) by (7) yields the timescale of peak discharge,
while the time for the field to reach this maximum is
$\sim l/c$.  The ratio of these two times is given by:
\begin{equation}
\frac{\tau_{dissipation}}{\tau_{formation}} = 29.0 n_{36} U_{34}^{-\frac{9}{8}}
f^{-\frac{3}{2}} g^{-1}
\end{equation}
and is $\geq 1$ for the parameter regime of concern.
The power of pair production at the FB surface,
$d \epsilon /dt$, is the product of $\frac{d^2 \epsilon}{dVdt}$ and
$4 \pi r^2 l$ where $r$ is the FB radius and $l$ is given by (5).
This yields:
\begin{equation}
\left[ \frac{d \epsilon}{dt} \right]_{pairs} =
1.8 \times 10^{55} n_{36}^{-1} r_{10}^2
U_{34}^{\frac{21}{8}} [f(U_{30})]^{\frac{7}{2}} g(U_{34})~~~~{\rm ergs}~
{\rm s}^{-1}
\end{equation}
where $r_{10}$ is $r$ in units of 10 km.

The rate at which electrostatic field energy develops to
its peak value at the FB surface is given by
the product $\frac{E^2}{8 \pi} \times 4 \pi r^2 l \times \frac{c}{l}$, and may
be expressed as:
\begin{equation}
\left[ \frac{d \epsilon}{dt} \right]_{field} = 5.1 \times 10^{56} r_{10}^2
U_{34}^{\frac{3}{2}} [f(U_{34})]^2~~~~{\rm ergs}~{\rm s}^{-1}
\end{equation}
where use has been made of (3) and (4).  The excess of (11) over (10) is by
the same ratio as that of the field discharge to formation times given in (9).
However, for an efficient conversion of the FB energy into
surface pairs, one not only must equalize the two rates (10) and (11) at peak
field, but this rate must also be 
$\geq$ the outward energy flux
of the explosion itself.  For the parameters of the initial FB
used here, the total energy flux across a given surface due to the
expanding FB is higher than the rates in (10)
and (11), and is given by:
\begin{equation}
\left[\frac{d \epsilon}{dt}\right]_{FB} = 4 \pi r^2 cU = 
3.8 \times 10^{57} r_{10}^2 U_{34}~~~{\rm ergs}~{\rm s}^{-1}
\end{equation}
In fact, when
parity between (11) and (12) is reached, we have the radius
independent relation:
\begin{equation}
\frac{1}{8 \pi} \left( \frac{\sigma_c U}{e} \right)^2 = U
\end{equation}
which requires an
energy density of the surface electric field (6) equal to
that of the FB.  For 
$U_{34} \sim 1$	the former falls short of the latter
by a factor $\sim 10$.  We shall, however, discuss two
possible scenarios (not exhaustive) under which the rates (10) and (11) may
be comparable, and be on par
with (12) at some early phase of the explosion.  The Schwinger
mechanism may then be important.

The first scenario maintains our default values of
$r_o$ and $U_o$, but takes advantage of the fact that in the foregoing
development
$U$ refers to the energy density of the FB
{\it apparent} to a
laboratory observer at rest w.r.t. the ambient medium, the only
exception being (3) and (4) where $U$ is the black body energy density at
the mean laboratory frequency of the radiation blast.
As the FB expands a little, and its
bulk flow develops a Lorentz factor $\gamma$, volume expansion
reduces the co-moving number densities of the FB by $1/\gamma^3$, while
adiabatic cooling scales (Meszaros, Laguna and
Rees 1993) the co-moving energy density as
$1/\gamma^4$ and the energy of a
typical photon as $1/\gamma$.
Upon transformation to the
laboratory system,
the length contraction
effect, relativistic beaming, 
and the `blueshift' of particle energies by a 
factor $\sim \gamma$ imply that the photon frequency,
number density, and energy
density
are increased from their corresponding
co-moving values by a factor of $\gamma$,
$\gamma^5$ and $\gamma^6$ respectively.
Thus, at finite $\gamma$ 
the effect of the FB on the ambient medium should
be calculated using number and energy densities which are higher than their
initial values by $\gamma^2$ times
(e.g. $U \sim \gamma^2 U_o$), except $\sigma_c$,
which  remains 
unaffected because the photon blueshift is counteracted by adiabatic losses.

The pair efficiency requirement (13) may then be met
at $\gamma \geq 3 [f(U_{34})]^{-1} U_{34}^{-\frac{1}{4}}$.  For a FB
with $U_o$ of $U_{34} \sim 1$ the expansion can easily reach
$\gamma \geq 3$ (at which
point $\geq$ 50 \% of $U$ is due to radiation).  The criterion on $\xi$ for
this to happen, $\xi \leq 8.8 \times 10^{-3} = \xi_o$, is consistent with
an outflow from neutron star merging (Lattimer and Schramm 1976), and
is much less stringent than that of canonical FB models, since
$\xi/\xi_c \sim$ 100.
On the
requirement of rapid discharge
the ratio in (9), now modified by the fact that 
$\sigma_c$ does not scale with $\gamma$, is $\sim$ 1 for
$n_{36} \geq 1$.  Under these conditions most of the FB energy 
$\epsilon_o$ will
be converted to pairs, and will eventually emerge as a GRB.
Since the 
initial FB baryon density
is $\geq 2.5 \times 10^{36} (\xi/\xi_o) r_{10}^{-3} cm^{-3}$, 
and the laboratory
density of the FB at the time of discharge is $\gamma^2$ times higher,
the density ratio across the active front is a
reasonable $\eta \sim$ a few \%.

In our second scenario an efficient discharge occurs at the initial
explosion $\gamma =$ 1.  A larger
value for
$U_o$ at the
trigger, $U_{34} \geq 20$, or 
$\epsilon_o \sim$ 9 $\times$ 10$^{53}$ ergs
for $r_o \sim$ 10 km, is assumed.
This will be a limit where (11) is slightly less than (12), so that even though
(10) can be as large as (11)
for $n_{36} \geq 2$, only $\sim$ 80 \% of $\epsilon_o$ is available to drive
a GRB.  Under optimal conditions,
therefore, the burst will have total energy $\sim$ 7 $\times$ 10$^{53}$ ergs,
which may be relevant to the more energetic events such as the one detected
recently (Kulkarni et al 1998)
There is no restriction at all on $\xi$; but
for $\xi \leq$ 1 the FB is radiation dominated and has energy
$< M_\odot c^2$.
The FB has number density of baryons
$\leq 3 \times 10^{38} \xi$ cm$^{-3}$ and $kT \geq$ 200 MeV, which
renders it fortuitously resembling the inner core of a supernova.
The density ratio
$\eta \geq 0.7/\xi$ \%.
%reasonable to assume that the
%environment just outside the front is cooler than the 100 MeV limit 
%necessary for
%the existence of ambient protons and electrons.

Since the region outside the created pairs is essentially the ambient medium
which is likely to be optically thin to the pairs, a conservative estimate
of the baryon contamination is obtained by assuming that all the baryons of
the original electric double-layer are dragged out after 
discharge.  Then the Entropy-per-baryon of the flow is $>$
$U/(n_i m_p c^2)$, and is $> 1/\eta$ times higher than that of the FB
itself.  The formula provided by 
Meszaros, Laguna and Rees (1993) shows that
the flow will reach a maximum speed given by
$\gamma > 2.4 \times 10^{-2} (\epsilon_o/4 \times 10^{52} ergs) \xi^{-1}
\eta^{-1}$.  In the first scenario where $\epsilon_o \sim$ 1 and
both $\eta$, $\xi$ are $\sim$ 1 \%,
$\gamma >$ a few $\times$ 10$^2$, 
well within the range of values capable of delivering   
a GRB at the terminal relativistic shock.  Similarly in
the second scenario where $\epsilon_o \sim$ 20 and $\eta \sim$ 1 \%, we have
$\gamma >$ 100 if $\xi \sim$ 0.5.

%highly relativistic expansion in the context of canonical FB models.  In fact
%the bulk Lorentz factor $\gamma$ of the FB outflow will reach maximum at
%a laboratory radius of only
%$r/r_o \sim \gamma = 2.4 \times 10^{-2} E_{FB} M_b/M_\odot \geq 2.72$,
%where $E_{FB} is in units of $E_o$ (Meszaros, Laguna \&
%Rees 1993) and the formula does not suffer from further complications due
%to opacity transitions because the FB remains optically thick at such early
%epochs.  

In conclusion, this {\it Letter}  demonstrates 
that a fundamental quantum
process which takes place at the surface of a powerful 
celestial explosion has hitherto been ignored.
The resulting pair outflow can have the right properties
to form
a GRB at large radii.
If the proposed mechanism is relevant,
further research on the details of its operational environment may
shed light on the origin of GRBs.

%\begin{figure*}[h]
%\vspace{2cm}
%\special{psfile=grb_func.ps hoffset=125 voffset=350 hscale=55 vscale=55 angle=90}
%\vspace{1cm}
%\caption{The function f(U) as defined in the text.}\label{fig2}
%\end{figure*}

We thank Martin Rees,
Bohdan Paczynski, Stan Woosley, John Doty and Walter Lewin
for helpful discussions.

\vspace{2mm}

\noindent
{\bf References}

\noindent
1.~Costa, E. et al, {\it Nature}, {\bf 387}, 783 (1997). \\
\noindent
2.~van Paradijs, J. et al, {\it Nature}, {\bf 386}, 686 (1997). \\
\noindent
3.~Frail, D.A., Kulkarni, S. R., Nicastro, L., Feroci, M.
\& Taylor, G. B.,
{\it Nature} {\bf 389}, 261 (1997). \\
\noindent
4.~Metzger, M.R. et al, {\it Nature}, {\bf 387}, 878 (1997). \\
\noindent
5.~Kulkarni, S.R. et al, {\it Nature}, {\bf 393}, 35 (1998). \\
\noindent
6.~Meszaros, P. \& Rees, M.J., {\it Astrophys. J.}, {\bf 405}, 278 (1993). \\
\noindent
7.~Paczynski, B., in {\it Ann. NY Acad. Sci.}, {\bf 688}, 321 (1993). \\
\noindent
8.~Quenby, J.J. \& Lieu, R., {\it Nature}, {342}, 654 (1989). \\
\noindent
9.~Meszaros, P. \& Rees, M.J., {\it Astrophys. J.}, {\bf 476}, 232 (1997). \\
\noindent
10.~Paczynski, B. \& Rhoads, J.E., {\it Astrophys. J.}, {\bf 418}, L5 
(1993). \\
\noindent
11.~Vietri, M., {\it Astrophys. J.}, {\bf 478}, L9 (1997). \\
%\noindent
%Ramaprakash, A.N. et al, {\it Nature}, {\bf 393}, 43 (1998). \\
\noindent
12.~Schwinger, J., {\it Phys. Rev.}, {\bf 82}, 664 (1951). \\
\noindent
13.~Meszaros, P., Laguna, P., \& Rees, M.J., {\it Astrophys. J.}, {\bf 415},
181 (1993). \\
\noindent
14.~Lattimer, J.M. \& Schramm, D.N., {\it Astrophys. J.}, {\bf 210},
549 (1976). \\

\vspace{2mm}

\noindent
{\bf Figure Caption}

\noindent
Figure 1.  {\it Left:} radial outflux of radiation induces a surface 
electric field (sometimes called a Pannekoek-Rosseland field) by
accelerating ambient electrons away from protons.  {\it Right:} if this
electric field exceeds $E_c$ its discharge will result in the creation
of e$^+$e$^-$ pairs outside the FB.  The pairs can thus expand without
encountering more FB matter.

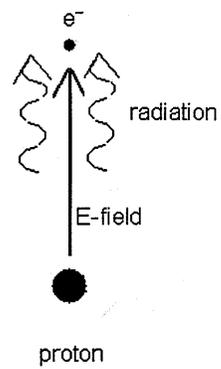
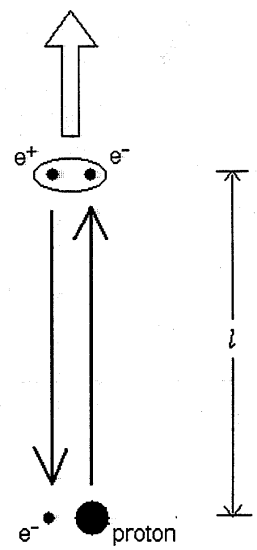
\end{document}